# Deep Learning for Super-resolution Ultrasound Imaging with Spatiotemporal Data

Arthur David Redfern [1], Katherine G. Brown [2]

[1] Department of Computer Science, University of Virgina, Richmod, Virginia, USA
[2] Department of Bioengineering, University of Texas at Dallas, Richardson, Texas, USA

*Abstract*—**Super-resolution ultrasound imaging (SRUS) is an active area of research as it brings up to a ten-fold improvement in the resolution of microvascular structures. The limitations to the clinical adoption of SRUS include long acquisition times and long image processing times. Both these limitations can be alleviated with deep learning approaches to the processing of SRUS images. Common approaches are based on the popular U-Net architecture while highly successful on medical image tasks was not optimized for microbubble (MB) contrast agent localization. Further, such architectures are focused only on localization of MB, and not on other aspects of the SRUS signal processing chain such as tissue signal rejection and MB detection. In this study we propose an optimized architecture based on modern improvements to convolutional neural networks from the ConvNeXt architecture and further customize the choice of features to improve performance on the specific tasks of both MB detection and localization within a single network. We employ a spatiotemporal input of up to five successive image frames to increase the number of MBs detected. The output structure produces three classifications: a MB detection Boolean for each pixel in the central image frame, as well as x and z offsets at 4-fold subpixel resolution for each MB detected. Ultrasound simulations generated images based on the L22-14v transducer (Verasonics) for training and testing of the proposed SRUS-ConvNeXt network.** *In vivo* **image data of a mouse brain was used as further validation of the architecture. The proposed network had the highest performance as measured by F1 score when configured for a 3-frame spatiotemporal input. The smallest localization error of $\lambda/22$ was achieved when the network was configured for a single input frame. The network prediction time was 1.78 ms on an NVIDIA 2080-Ti GPU. The flexibility of the proposed architecture allows extension to 10-fold upscaling for SRUS images with a much lower impact to number of parameters and subsequent increase in inference time than typical U-Net style approaches. This network is promising in the quest to develop a SRUS deep network architecture for real time image formation.**

## I. INTRODUCTION

Super-resolution ultrasound (SRUS) has demonstrated tremendous potential for use in revealing microvascular details beyond the reach of Doppler flow imaging and allowing better understanding of both the disease state and disease progression in cancer and diabetes, as well as improving the clarity of functional imaging of the brain [1]–[3]. SRUS provides up to a 10-fold improvement in resolution of the smallest vessels in the body when compared to conventional diffraction limited contrast-enhanced ultrasound imaging (CEUS) [4]. To accomplish this, a microbubble (MB) contrast agent is detected across thousands of image frames and the center of each MB estimated on up to a 10-fold finer grid. Clinical translation has been limited by the requirements for long acquisition times, the difficulty of correcting image frames for patient motion, and extended computation time. Deep learning techniques have been shown to reduce computation times by factors of more than 10, and to reduce acquisition times by achieving robustness in the situation of overlapping MB signals. [5]–[7].

This study was supported by NIH grants R01EB025841 and R01DK126833 and Texas CPRIT RP180670.

Deep learning for SRUS localization has most often been accomplished with variations of the highly successful U-Net architecture of Ronneberger *et al.*[8]. U-Net has been effectively used in medical image segmentation beyond the original use to identify neuronal structures in electron microscopy, to brain tumor detection, retinal vessel segmentation, and many other applications. Even so, the architecture is not the most efficient for SRUS in that an upscaled image is produced by inverted convolutions which grow exponentially as the upsampling factor is increased. Further, the U-Net downsampling architecture makes the assumption that the scale of the target, here a MB, varies, when in fact the PSF of MB is quite static in plane wave imaging, and undergoes minor changes with focused imaging. These factors can be taken advantage of by use of a classification approach instead of a segmentation approach for the head end of the network, removing the encoding U-Net structure, as shown in our prior work [6]. In addition, the decoding structure of the U-Net architecture can be reduced in levels and the kernel sizes can be fixed due to the fixed size of the MB PSF.

The transformer network architecture using an attention mechanism which improved the performance of language models has been the subject of much experimentation in the imaging field. Liu *et al.* examined the improvements in performance of this architecture and concluded that much of the gains were from advanced training methodologies rather than due to the attention mechanism. They proposed a convolution network having the best elements of new training methods in ConvNeXt and showed that it outperformed the then most advanced attention based networks for imaging [9]. We adopt some of these core principles here in our proposed architecture SRUS-ConvNeXt.

The use of spatiotemporal data contains information of the motion of MB in the field of view and we hypothesize that this could be used by a network to enhance the detection of MBs, potentially reducing the acquisition time. Multiple frames of input data will increase the size of the network, and thus increase the inference time, so a balance will be required.

In this study we introduce a new deep learning network architecture optimized for detection and localization of MB in SRUS and examine the effect of the length of image stack input to on the detection rate and the localization accuracy of MB by the network.

## II. MATERIALS AND METHODS

*A. Network Architecture:*

The network architecture had input of 228 x 228 x *m* where *m* can be any integer, and *m* = 1, 3, and 5 images in this study. The output of the network consisted of 228 x 228 x 3 where each pixel of the input image was mapped to 3 outputs, the first a Boolean indicating the presence/absence of a MB center, and the other two indicating the x and z sub-pixel offsets of the location of the MB on a 4-fold finer grid.

A building block based on ConvNeXt featuring depthwise separable convolutions and inverted residuals for efficiency and including an identity path was featured in the architecture which had no down or upsampling, see Fig. 1. The building block had three 2-D convolution layers, each followed by batch normalization and Gaussian Error Linear Unit (GELU) nonlinearity. The first convolution was a 7 x 7 grouped convolution of 128 channels and the other two 1 x 1 convolutions with a maximum dimension of 768 channels. The block was repeated 6 times and preceded by a 2D convolutional layer using a 3 x 3 kernel with its input being the network input and output having 128 channels. And it was followed by a parallel decoder architecture using 1 x 1 convolutions to form three decoders; one for microbubble detection and the other two for x and z offset prediction.

The loss term for training was the weighted sum of three components; the MB detection loss, the x offset loss and the z offset loss. The x and z offset losses were computed only in places where a MB was present. The weightings used were 0.9 for MB detection and 0.1 each for x and z offsets. Training used stochastic gradient descent (SGD) with momentum and L2 regularization for the weight update and proceeded for 50 epochs. The network was implemented in custom PyTorch code.

*B. Simulations:*

A custom simulation was developed in MATLAB software (MathWorks Inc, Natick, MA) around the SIMUS simulator in the MUST toolbox [10] to generate synthetic US images based on the transducer parameters of the L22-14v (Verasonics, Kirkland, WA) with a variable signal-to-noise ratio (SNR, 1.5 to 5.2 dB). MB flowed from frame to frame at a variable rate in vessels of small but varying diameter (100 to 500 μm) as shown in Fig. 2. The tissue background was formed by the introduction of randomly placed scatterers. The MBs were formed by moving scatterers with a relative reflectivity given by the selected SNR level. A total of 12,000 training images at three SNR levels were generated using a combination of simulation and image augmentation by horizontal flipping and cropping. Similarly, an additional 5 sets of 2,000 images each at three SNR levels were generated separately for testing using a different vessel morphology.

*C. Imaging:*

*In vivo* US imaging was performed in anesthetized animals after administering a 12.5 μL bolus injection of a MB contrast agent (Definity, Lantheus Medical Imaging, N. Billerica, MA) diluted in 12.5 μL saline and delivered via a tail vein catheter. Imaging was performed with a Vantage 256 (Verasonics) equipped with an L22-14v transducer. Angular compounding of 5 angles between -10 and 10 degrees was implemented resulting in an overall frame rate of 500 Hz. The transmit center frequency was 15 MHz. RF data was captured and stored offline for analysis. A total of 36,000 images were collected for *in vivo* testing of the proposed network.

*D. SRUS Processing:*

The RF data collected off-line was beamformed with the Verasonics built-in beamforming resulting in IQ data. Pre-processing of IQ data for tissue signal and noise reduction was performed, along with envelope detection, prior to MB detection and localization with the deep network.

The tissue signal was suppressed using block-wise SVD filtering. Based on slow time analysis, the lower cutoff for filtering for each block was obtained by finding the singular value at which the slope of the singular value curve was 30º with the abscissa [11]. A fixed singular value of 400 was the maximum singular value included in the filtered result to reduce noise.

Image frames were presented to the deep network in groups of 1, 3 or 5 consecutive frames using a sliding window approach. The network output represented a prediction for the MB localizations for the center frame of the input stack. Localized values were accumulated across all frames produced by the deep network to form the final SRUS image. Microvascular structures were enhanced by a modified Hessian-based filter [12], [13].

*E. Evaluation:*

After training the three versions of the network for the three configurations of input stack sizes, each network was used to predict MB localizations for the 5 test sets at each of 3 SNR levels. Performance was measured by localization accuracy as well as F1 score and results presented as mean +/- standard deviation.

F1 score was calculated as:

$$F1 = \frac{2 * Precision * Recall}{Precision + Recall}$$

Where

$$Recall = \frac{TP}{TP + FN}$$

$$Precision = \frac{TP}{TP + FP}$$

And TP are True Positives, FP are False Positives, and FN are False Negatives.

## III. RESULTS

The SRUS-ConvNeXt network had the lowest localization error of 4.68 +/- 0.23 μm configured with an input image stack of 1 image and having a small degradation in the average localization error for input image stacks of 3 and 5 images respectively, as depicted in the bar chart of Fig. 3. This represents an error of approximately λ/22 given the acoustic wavelength λ at 15 MHz of 103 μm. As expected, with increasing SNR the network localization performance showed improvement.

In contrast, the SRUS-ConvNeXt network performed best as measured by an F1 score (72.02 +/- 0.08 %) with the spatiotemporal input stack of 3 images, although this advantage was slight over the 5-input image network, as shown in Fig. 4. The SNR level of the test images was seen to have a minimal impact on the F1 score.

Testing the network on *in vivo* data from a mouse brain revealed a high definition image with similar levels of detail to those formed by conventional SRUS methods. The SRUS image based on a configuration of a 3 frame input stack is depicted in Fig. 5.

The size of the network was 838,540 parameters. The network took an average of 1.78 ms per 228 x 228 pixel image to predict a single high resolution SRUS image frame on an NVIDIA 2080-Ti GPU.

## IV. DISCUSSION

A novel approach for SRUS image formation using an optimized deep network for MB detection and localization was introduced. SRUS-ConvNeXt efficiently predicted SRUS images with low localization error and having a level of detail consistent with conventional methods.

Deep networks have been studied for SRUS localization as they represent an efficient solution to a computationally intensive processing step. They have been shown to have other attractive properties such as the ability to tolerate overlapping MB signals from dense MB injections, and to have small localization errors.

In this study, the average localization error of MB at a high concentration was λ/22, which is comparable to that previously demonstrated of ~λ/12 by Van Sloun *et al.* [14] and λ/29 by Youn *et al.* [15]. Use of more frames of spatiotemporal input improved F1 score, specifically by increasing the number of MB detected. However, the localization error decreased by a small amount. In future work we will adjust architecture parameters in the network to improve localization error as the higher MB detection rate has the benefit of reducing image acquisition time. The demonstrated average frame processing time of 1.78 ms per 228 x 228 pixels represents a 10-fold improvement over conventional SRUS and is promising for operation in real time.

Additionally, the architecture demonstrates improved scalability in performance over fully convolutional U-Net style networks due to its unique architecture of paired axial and lateral offsets for classification rather than upsampling over the entire image. The architecture flexibility allows extension to 10-fold upscaling for SRUS images with a much lower impact to number of parameters and subsequent increase in inference time than typical U-Net style approaches. The results are promising for the extension of this architecture to one achieves 100-times resolution (10x upsampling) while reducing both acquisition time and localization time.

## V. CONCLUSIONS

A novel automated approach for producing SRUS images was introduced and then evaluated in both simulation and with *in vivo* experiments. Use of the proposed SRUS-ConvNeXt deep network shows potential for improving performance of SR-US towards a real-time imaging modality scalable to 10-fold upsampling.

COMPLIANCE WITH ETHICAL STANDARDS

Animal experiments were approved by the Institutional Animal Care and Use Committee (IACUC) at the University of Texas at Dallas.

CONFLICTS OF INTEREST

The authors have no conflicts of interest to disclose.

ACKNOWLEDGEMENTS

The authors appreciate the initial work on this project by Scott Chase Waggener, the many helpful conversations with Dr. Arthur D. Redfern, and the contributions to MB simulations of Mia Sargusingh. This work was accomplished with the support of the University of Texas at Dallas laboratories and equipment.

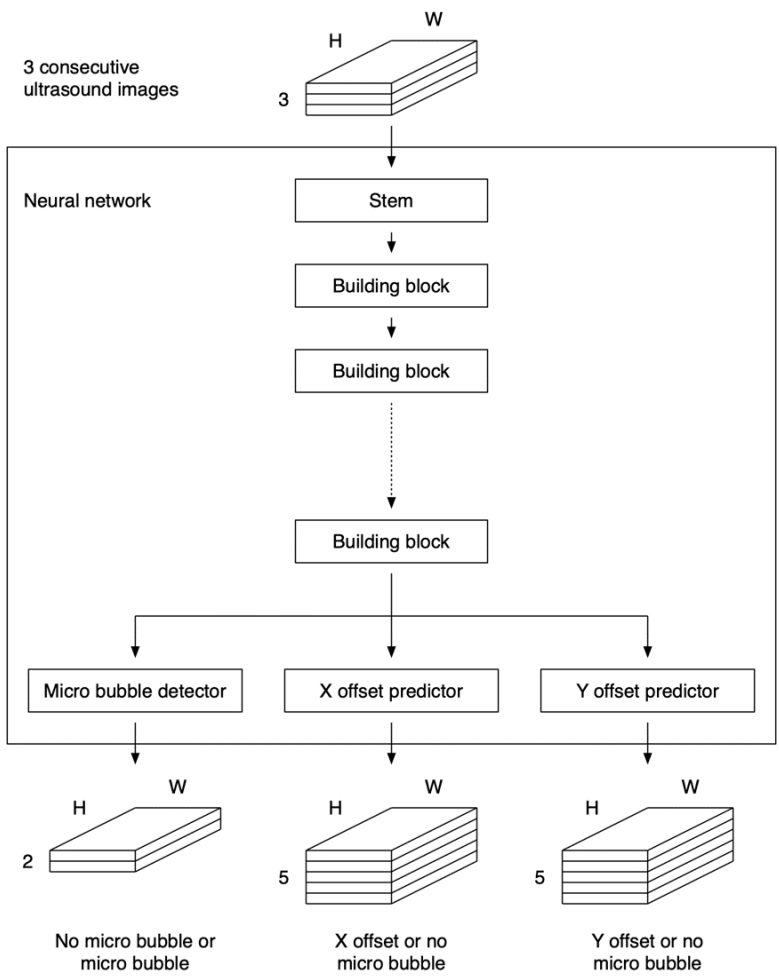

Fig. 1. SRUS-ConvNeXt architecture showing a 3-image sequence spatiotemporal input and three parallel decoders for high resolution output.

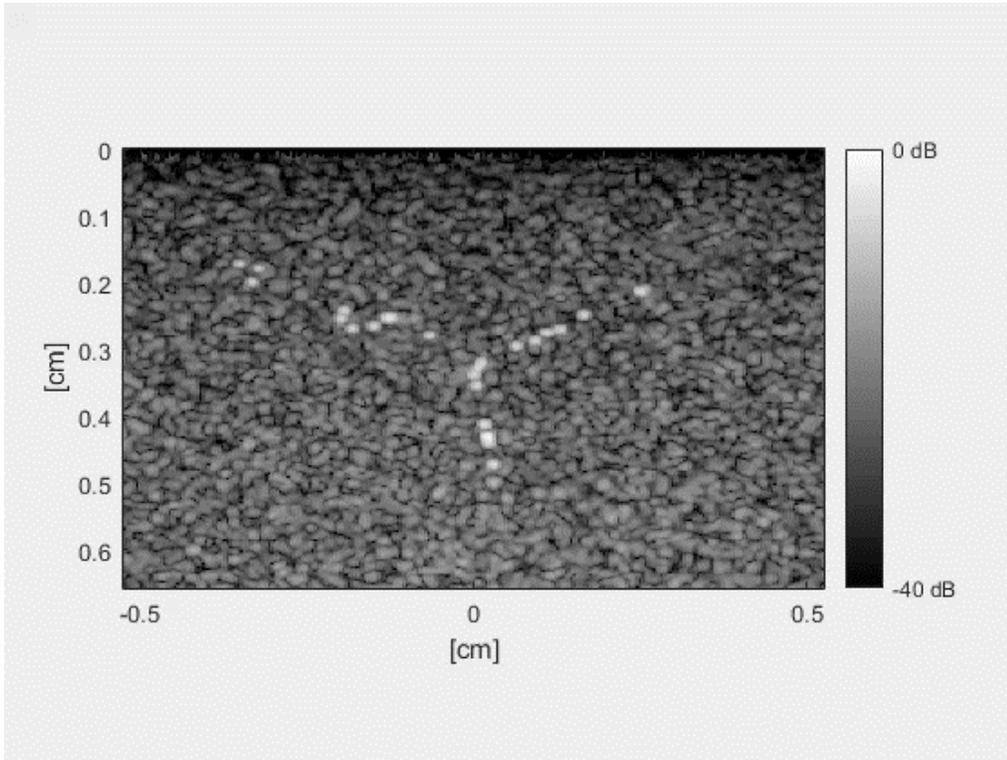

Fig. 2. Simulated ultrasound image based on a Verasonics L22-14v transducer at 15 MHz of microbubbles (MBs) moving in a bifurcating vessel across a tissue background with a signal to noise ratio (SNR) of 1.6 dB.

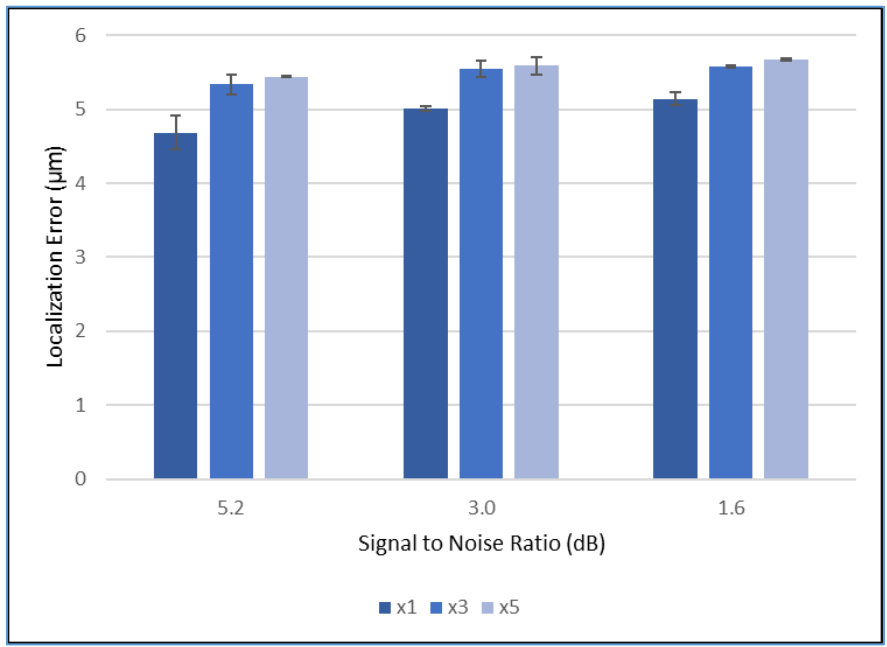

Fig. 3. Localization error shown for test datasets for spatiotemporal inputs of 1 image frame (x1), 3 image frames (x3), and 5 image frames (x5). The localization error is minimized for the x1 case. Error bars represent the standard deviation.

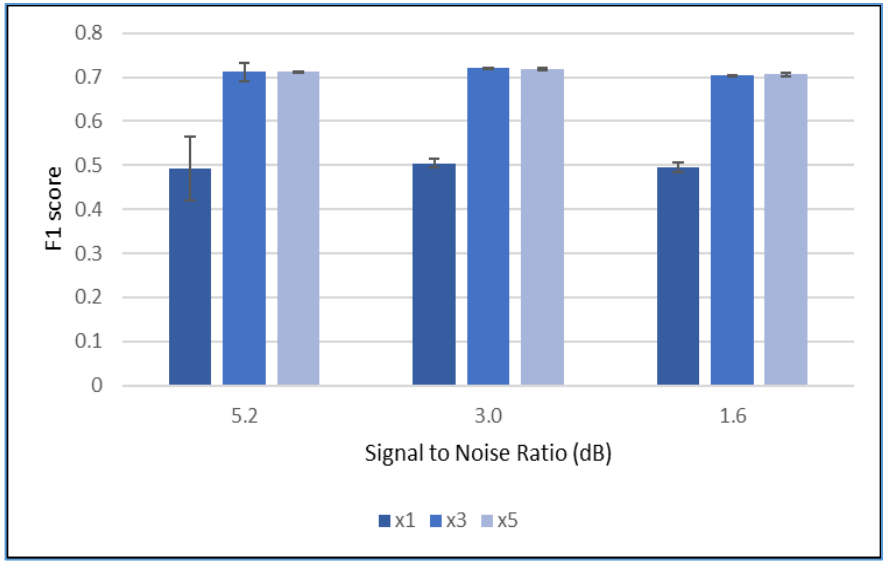

Fig. 4. F1 score shown for test datasets for spatiotemporal inputs of 1 image frame (x1), 3 image frames (x3), and 5 image frames (x5). F1 score is maximized for the x3 case. Error bars represent the standard deviation.

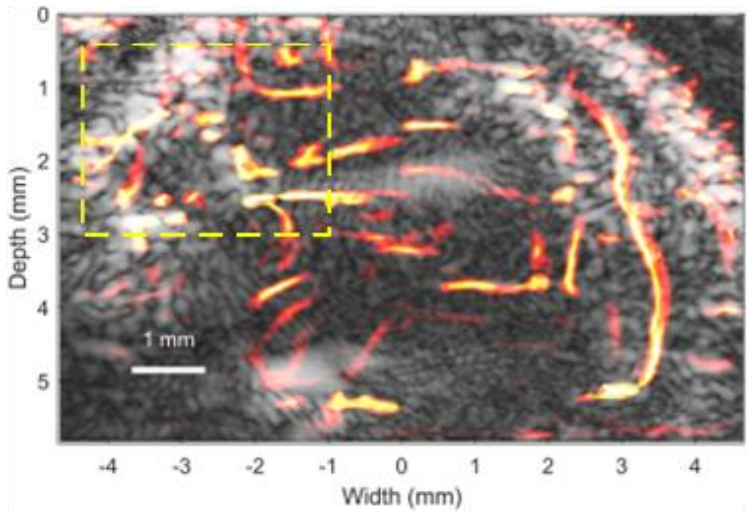 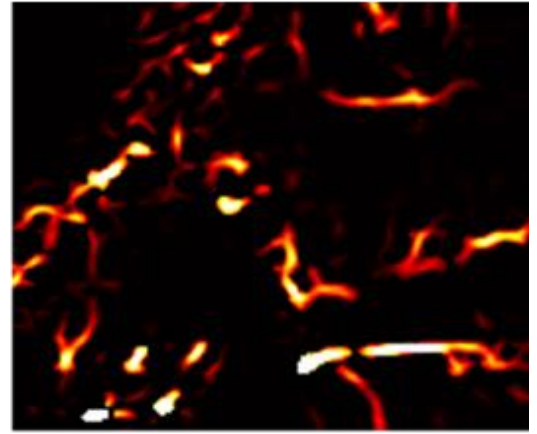

Fig. 5. Predicted SRUS image from the proposed deep network with spatiotemporal input of 3 image frames of *in vivo* data from mouse brain images through an intact skull using 36,000 frames collected with Verasonics Vantage 256, L22-14v transducer, 5 angles compounded, and 500 Hz frame rate. The inset image delineated by a dashed yellow box on left is enlarged on the right.